\documentclass{ifacconf}

\usepackage{graphicx}      
\usepackage{natbib}        
\usepackage{amsmath, amssymb}
\usepackage{graphicx}
\usepackage{tikz}
\usetikzlibrary{matrix,arrows.meta}

\newcommand{\ov}{\overline}
\newcommand{\be}{\begin{equation}}
\newcommand{\ee}{\end{equation}}
\newcommand{\mc}{\mathcal}
\newcommand{\R}{\mathbb{R}}
\begin{document}
\begin{frontmatter}

\title{Graphical Games and Decomposition\thanksref{footnoteinfo}} 

\thanks[footnoteinfo]{This work was partially supported by MIUR grant Dipartimenti di Eccellenza 2018--2022 [CUP: E11G18000350001].}

\author[First]{Laura Arditti} 
\author[First]{Giacomo Como}
\author[First]{Fabio Fagnani}

\address[First]{{Department of Mathematical Sciences, Politecnico di Torino}\\ Corso Duca degli Abruzzi 24, 10129 Torino, Italy\\ 
(e-mail: \{laura.arditti,giacomo.como,fabio.fagnani\}@polito.it).}

\begin{abstract}                
%
	We consider graphical games as introduced by \cite{Kearns.ea:2001}. First we analyse the interaction of graphicality with a notion of strategic equivalence of games, providing a minimal complexity graphical description for games. Then we study the interplay between graphicality and the classical decomposition of games proposed by \cite{Candogan.ea:2011}, characterizing the graphical properties of each part of the decomposition.
\end{abstract}


\end{frontmatter}

\section{Introduction}
Graphical games, which were first introduced in \cite{Kearns.ea:2001}, are games equipped with a network structure among players that specifies the pattern of dependence of their utilities. More precisely, in a graphical game the utility of a player is made to depend only on her own action and the action of her out-neighbour players. 
They have recently become prominent as a unifying theory to study the emergence of global phenomena in socio-economic networks like peer effects, technology adoption, and consensus formation (\cite{Jackson.Zenou:2015}). They are also a natural model in engineering and computer science to describe the interactions in multi-agent systems and a powerful tool to design distributed algorithms (\cite{Daskalakis.Papadimitriou:2006}).

While there is already a large amount of literature focusing on specific graphical games (e.g. coordination and anti-coordination games), it is still missing a general theory. 
How the graphicality of a game reflects on its properties is still largely unexplored. A remarkable exception is constituted by the paper by \cite{Babichenko.Tamuz:2016} where authors prove that a potential graphical game admits a potential reflecting the graphical structure, that can be decomposed as a sum of terms defined on the cliques of the graph. 
For non potential games, at the best of our knowledge, there is no such general analysis. 

In this paper we focus on a concept of strategic equivalence for games and on a classical decomposition of games in terms of non-strategic, potential and harmonic parts introduced in \cite{Candogan.ea:2011}. An analysis of some strategical aspects of such decomposition has been carried out in \cite{Scarsini.ea:2011} where authors point out some drawback of it and propose a generalization to overcome those issues.
In our work, instead, we undertake a fundamental analysis on how such concepts interact with the graphicality of a game. 

In many contexts, it is natural to consider games up to strategic equivalence, meaning that we are only concerned with variations of the utility of a player when she modifies her action rather then their absolute values. This is typical in technological contexts where the game is the result of an explicit design rather then an intrinsic model. Classical evolutionary dynamics associated to games like the best-response dynamics or the log-likelihood dynamics are invariant with respect to this equivalence. Our first result, Corollary \ref{corollary:minimal-class}, determines the minimal graph with respect to which a game is graphical up to strategic equivalence. It is a sort of minimal complexity result showing that certain interactions in the game are fictitious and can consequently be removed. 

Our main results regard the way the decomposition reflects the graphicality of a game. While the non-strategic part is always graphical with respect to the same graph than the original game, the same is not true in general for the potential and harmonic parts that are instead graphical with respect to a larger graph where out-neighbourhoods have become cliques.
Intuitively, this means that there are short range hidden strategic interactions which involve only players that directly influence the utility of some common player in $\mc G$.
This is the content of our main result, Theorem \ref{theo:main}. This ``interaction enlargement'' not always happens and we show that for the important class of pairwise-separable graphical games (where utility of players is the sum of utilities of $2$-players games played with their neighbours) actually the potential and harmonic components maintain the original graphicality on $\mc G$. 
We present an explicit example where this phenomenon shows up.
%

\section{Graphical games and strategic equivalence}

\subsection{Graph-theoretic notation} 

A (directed) graph $\mc G=(\mc V,\mc E)$ is the pair of a finite node set $\mc V$ and a link set $\mc E\subseteq\mc V\times\mc V$, whereby a link $(i,j)\in\mc E$ is meant as directed from its tail node $i$ to its head node $j$. We shall assume that the graph $\mc G$ contains no selfloops, i.e., that $(i,i)\notin\mc E$ for every $i\in\mc V$ and we shall denote by $\mc N_{i}=\{j\in\mc V:\,(i,j)\in\mc E\}$ and $\mc N^{\bullet}_i=\mc N_{i}\cup\{i\}$ the open and, respectively, closed out-neighbourhoods of a node $i$ in $\mc G$. The intersection of two graphs $\mc G_1=(\mc V,\mc E_1)$ and $\mc G_2=(\mc V,\mc E_2)$ is the graph $\mc G_1\cap\mc G_2=(\mc V,\mc E)$ where $\mc E=\mc E_1\cap\mc E_2$. Note that we shall consider undirected graphs as a special case where $(i,j)\in\mc E$ if and only of $(j,i)\in\mc E$. 

We shall often consider certain supergraphs of a graph $\mc G=(\mc V, \mc E)$, all undirected and obtained by keeping the same node set $\mc V$ and augmenting the link set $\mc E$ as follows. First, let $\mc G^{\leftrightarrow}=(\mc V,\mc E^{\leftrightarrow})$ where 
$$\mc E^{\leftrightarrow}=\mc E\cup\left\{(j,i):\,(i,j)\in\mc E\right\}$$ 
be the graph obtained by making all links in $\mc G$ undirected. On the other hand, let $\mc G^{\triangle}=(\mathcal{V}, \mathcal{E}^{\triangle})$ be the graph whose link set $\mathcal{E}^{\triangle}$ is obtained from $\mc E$ by adding links among all pair of outneighbours of every node, i.e., 
$$\mathcal{E}^{\triangle}=\mc E^{\leftrightarrow}\cup\bigcup\limits_{i\in\mc V}\left\{(j,l):\,j\ne l\in \mc N_{i}\right\}\,.$$
In fact, we shall consider a more general notion of extension of a graph $\mc G=(\mc V, \mc E)$ defined as follows. Consider a covering
\be\label{eq:partitionS} \mc N_{i}=\bigcup_{h=1}^{k_i}\mc S^h_i \ee 
of the open neighbourhood of each node $i\in\mc V$. We denote the array of such coverings as $\mc S=\{\mc S^h_i\}_{i\in\mc V,1\le h\le k_i}$, and we call $\mc S$ a \emph{splitting} of $\mc G$. Then, define the graph $\mc G^{\mc S}=(\mc V,\mc E^{\mc S})$ with the same node set as $\mc G$ and link set 
$$\mc E^{\mc S}=\mc E^{\leftrightarrow}\cup\bigcup_{i\in\mc V}\bigcup_{h=1}^{k_i}\left\{(j,l):\,j\ne l\in\mc S_i^h\right\}\,.$$
Observe that $\mc G^{\leftrightarrow}$ and $\mc G^{\triangle}$ are special cases of $\mc G^{\mc S}$ when, for every node $i\in\mc V$, $k_i=|\mc N_{i}|$ and $\mc S_i^h$ are all singletons, and respectively, when  $k_i=1$ and $\mc S^1_i=\mc N_{i}$.

\subsection{Graphical games}
Throughout the paper, we shall consider strategic form games with a finite set of players $\mc V$, where each player $i \in \mathcal{V}$ has finite action set $\mc A_i$.  We shall denote by $\mathcal{X}= \prod_{i \in \mathcal{V}} \mc A_i$ the space of all players' strategy profiles. For a player $i\in\mc V$, let $\mathcal{X}_{-i}= \prod_{j \in \mathcal{V}\setminus\{i\}} \mc A_j$ be the set of strategy profiles of all players except for $i$. 
As customary, for a strategy profile $x\in\mc X$, the strategy profile of all players except for $i$ is denoted by $x_{-i}\in\mc X_{-i}$. When two strategy profiles $x,y\in\mc X$ coincide except for possibly in their $i$-th entry, i.e., when $x_{-i}=y_{-i}$, we shall say that $x$ and $y$ are $i$-comparable and we shall write $x\sim_i y$.  

Let each player $i \in \mathcal{V}$ be equipped with a utility function $u_i:\mathcal{X}\to \mathbb{R}$. We shall identify a game with player set $\mc V$ and strategy profile space $\mc X$ with the vector $u$ assembling all the players' utilities. Notice that, in this way, the set of all games with player set $\mc V$ and strategy profile space $\mc X$, to be denoted by $\Gamma$, is isomorphic to the vector space $\R^{\mc V\times\mc X}$. 

Graphical games (\cite{Kearns.ea:2001}) are defined imposing suitable restrictions on the way utilities depend on strategies. Precisely, a game  $u$ is said to be \emph{graphical} on a graph $\mc G = (\mathcal{V}, \mathcal{E})$ (or, briefly, a $\mc G$-\emph{game}) if 
the utility of each player $i\in\mc V$ depends only on her own action and the actions of fellow players in her  neighbourhood in $\mc G$, i.e., if 
\be\label{eq:graphical-def}u_i(x)=u_i(y)\,,\qquad\forall x,y\in\mc X\text{ s.t.~}x_{\mc N_{i}}=y_{\mc N_{i}}\,.\ee
Notice that if a game $u$ is graphical with respect to two graphs $\mc G_1 = (\mathcal{V}, \mathcal{E}_1)$ and $\mathcal{G}_2 = (\mathcal{V}, \mc{E}_2)$, it is also graphical with respect to $\mc G_1\cap\mc G_2$. Considering that every game $u$ is trivially graphical on the complete graph on $\mc V$, we can conclude that to each game $u\in\Gamma$ one can always associate the smallest graph on which $u$ is graphical. We shall refer to such graph as the \emph{minimal graph} of the game $u$ and denote it as $\mc G_u$.

A special case of graphical games are the \emph{pairwise-separable graphical games} (cf.~\cite{Daskalakis.Papadimitriou:2009,Cai.Daskalakis:2011}) whereby, for a given graph $\mc G=(\mc V,\mc E)$, the 
utility of player $i\in\mc V$ is in the form 
\begin{equation}\label{pairwise-game}
u_i(x) = \sum_{j \in \mc N_{i}} u_{ij}(x_i,x_j) \qquad \forall x \in \mathcal{X}\,,
\end{equation}
where $u_{ij}:\mc A_i\times\mc A_j\to\R$ for $i,j\in\mc E$. 
Notice that the utilities in \eqref{pairwise-game} clearly define a $\mc G$-game.
In fact, such a game can be interpreted as one in which the players are located at the nodes of $\mc G$ whose links are to be interpreted as two-player games between their endpoints (with the convention that $u_{ji}(x_j,x_i)=0$ for every $(j,i)\notin\mc E$ such that $(i,j)\in\mc E$). Every player $i\in\mc V$ can chose a unique action $x_i\in\mc A_i$ to be used in all games she simultaneously participates in and gets a utility that is the aggregate of the utilities from all her outgoing links.

Every graphical game possesses some separability properties, pairwise separability being the finest possible case. It is then useful to treat in a unified way all graphical games by introducing the notion of $\mc S$-separability.
More precisely, given a $\mc G$-game $u$ and a splitting $\mc S=\{S_i^h\}_{i \in \mc V, 1 \leq h \leq k_i}$ of $\mc G$, we say that the game $u$ is  $\mc S$-\emph{separable} if for every player $i \in \mc V$ the utility function can be decomposed as 
\begin{equation}\label{eq:utility-decomposition}u_i=u^0_i+\sum_{h=1}^{k_i}u_{i}^h\,,
\end{equation}
where $u_i^0$ depends only on the actions of players in $\mc N_i$, i.e., $u_i^0(x)=u_i^0(y)$ if $x_{\mc N_i}=y_{\mc N_i}$ and each term $u_i^h$ depends only on the action of $i$ itself and of players in $S_i^h$, i.e., $u_i^h(x)=u_i^h(y)$ if $x_i = y_i$ and $x_{S_i^h}=y_{S_i^h}$.
Separable graphical games are an extension of pairwise-separable graphical games, where we consider that players have separate interactions with different \emph{groups} of their neighbours. Such partition of their neighbourhoods is described by the splitting $\mc S$.

\subsection{Strategic equivalence and graphicality}

A game $u$ is referred to as \emph{non-strategic} if the utility of each player $i\in\mc V$ does not depend on her own action, i.e., if
\be\label{eq:non-strategic}	u_i(x)=u_i(y)\,,\quad\forall x,y\in\mc X\text{ s.t.~}y\sim_i x\,.\ee
The set of non-strategic games will be denoted by $\mc N$. 
Two games $u$ and $\tilde u$ are referred to as strategically equivalent if their difference is a non-strategic game, i.e., if 
\be\label{eq:strategically-equivalent}	u_i(x)-\tilde u_i(x)=u_i(y)-\tilde u_i(y)\,,\quad\forall x,y\in\mc X\text{ s.t.~}y\sim_i x\,.\ee
Strategic equivalence is in fact an equivalence relation on games and we shall denote the strategic equivalence class of a game $u$ by $[u]$. It follows from the definition that strategically equivalent games are separable with respect to the same splittings.

A game  $u$ is referred to as normalized if 
\be\label{eq:normalized}
\sum_{y\sim_i x} u_i(y) = 0\,,\quad \forall x \in \mathcal{X}, i\in\mc V\,.\ee
For a game $u$, one can define its normalized version as the game $\ov u$ with utilities 
\be\label{eq:u-normalized}
\ov u_i(x)=u_i(x)-\frac1{|\mc X_{-i}|}\sum_{y\sim_i x} u_i(y)\,,\quad \forall x \in \mathcal{X}, i\in\mc V \,.\ee
It is then easily verified that the game $\ov u$ is both normalized and strategically equivalent to $u$. In fact, $\ov u$ is the unique normalized game in the class $[u]$ \cite[Lemma 4.6.]{Candogan.ea:2011}.


Two strategically equivalent games $u$ and $\tilde u$ might have quite different minimal graphs $\mc G_u$ and $\mc G_{\tilde u}$. Indeed, it is easy to see that every game $u$ admits a strategically equivalent game $\tilde u$ such that $\mc G_{\tilde u}$ is the complete graph. It is then less obvious that a strategic equivalence class $[u]$ always contains a game whose minimal graph is contained in the minimal graph of every other game $\tilde u$ in $[u]$. This property turns out to hold true, as a consequence of the following result. 
\begin{prop}\label{prop:minimal-class}	
	Let $u$ be a game, $\ov u$ its normalized version. Then, 
	\be\label{eqGusubsetGu}\mc G_{\ov u}\subseteq\mc G_{u}\,.\ee
\end{prop}

The following is a direct consequence of Proposition \ref{prop:minimal-class} and of the fact that the normalized version of any strategically equivalent game $\tilde u\in[u]$ coincides with the normalized version $\ov u$ of $u$. 
\begin{cor}\label{corollary:minimal-class} 
	Let $u$ be a game, $\ov u$ its normalized version. Then, 
	\be\label{eqGusubsetGtildeu}\mc G_{\ov u}\subseteq\mc G_{\tilde u}\,.\ee
	for every strategically equivalent game $\tilde u\in[u]$. 
\end{cor}

Corollary \ref{corollary:minimal-class} implies that there exists a game in $[u]$ that is graphical on the intersection 
$$\mc G_{[u]}=\bigcap_{\tilde u\in[u]}\mc G_{\tilde u}$$ of the minimal graphs of all the strategically equivalent games $[u]$. 
In fact, the graph
$$\mc G_{[u]}=\mc G_{\ov u}$$
may be interpreted as the minimal topological complexity needed to represent a game in the class $[u]$, namely a game up to non-strategic equivalence. 



\section{Decomposition of graphical games}

We now introduce the two classes of potential games (\cite{Monderer.Shapley:1996}) and harmonic games (\cite{Candogan.ea:2011}). 
A game $u\in\Gamma$ is as an (exact) \emph{potential game} if there exists a function $\phi:\mathcal{X}\to\mathbb{R}$ such that 
\begin{equation}\label{eq:potential}u_i(x)-u_i(y)=\phi(x)-\phi(y)\,,\quad \forall i\in\mc V\,,\ \forall x\sim_i y\in\mc X\,,
\end{equation}
%
while $u \in \Gamma$ is an \emph{harmonic game} if 	
\begin{equation}\label{eq:harmonic}
\sum_{i \in \mathcal{V}} \sum_{y\sim_i x} [u_i(x) - u_i(y)] = 0\,,\quad \forall x\in\mc X\,.
\end{equation}
Notice that a normalized game is harmonic if and only if 
\begin{equation}\label{eq:harmonic-normalized}
\sum_{i \in \mathcal{V}}|\mc A_i| u_i(x)= 0\,,\quad \forall x\in\mc X\,.
\end{equation}
Hence, in particular, if the action sets of all players have the same cardinality $|\mc A_i|=a$, then a normalized game $u$ is harmonic if and only if 
it is a $0$-\emph{sum game}, i.e., 
\begin{equation}\label{eq:zero-sum}
\sum_{i \in \mathcal{V}} u_i(x)= 0\,,\qquad \forall x\in\mc X\,.
\end{equation}

\cite{Babichenko.Tamuz:2016} showed that the potential of a potential $\mc G$-game possesses a special structure that reflects the one of the underlying graph $\mc G$.
Corollary \ref{corollary:minimal-class} allows us to improve such result by refining the decomposition proposed therein.
Indeed, all strategically equivalent games to a potential game $u$ are still potential with the same potential function. So the graphical structure of the potential is determined by the \emph{smallest} graph $\mc G_{[u]}$.

\begin{cor}\label{corollay:graphical-potential}
	Let $u$ be a potential $\mc G$-game with potential $\phi$. Then $\phi$ can be decomposed as
	\begin{equation}
	\phi(x) = \sum_{C \in \mc C(\mc G_{[u]})} \phi^C(x_C)
	\end{equation}
	for some functions $\phi^C: \Pi_{i \in C} \mc A_i \rightarrow \mathbb{R}$ called \emph{local potentials} and where $\mc C(\mc G_{[u]})$ is the set of maximal cliques of $\mc G_{[u]}$.
\end{cor}

It has been proven in \cite[Theorem 4.1.]{Candogan.ea:2011} that the space of games can be decomposed as a direct sum  $\mathcal{P} \oplus \mathcal{N} \oplus \mathcal{H}$, where $\mathcal{P}$ is the space of normalized potential games, $\mathcal{N}$ is the space of non-strategic games, $\mathcal{H}$ is the space of normalized harmonic games. We are interested in characterizing graphicality properties of such decomposition. In particular, our main result, stated as Theorem \ref{theo:main} below, provides a generalization and refinement of the one in \cite{Candogan.ea:2011}. 

\begin{thm}\label{theo:main}
	A finite game $u\in\Gamma$ is $\mc S$-separable if and only if it can be decomposed as  
	$$u=u_{\mathcal{N}}+u_{\mathcal{P}}+u_{\mathcal{H}}\,,$$ 
	where
	\begin{itemize}
		\item $u_{\mathcal{N}}$ is a non-strategic $\mc G_u$-game; 
		\item $u_{\mathcal{P}}$ is a normalized $\mc S$-separable potential $\mc G_{[u]}^{\mc S}$-game; 
		\item $u_{\mathcal{H}}$ is a normalized $\mc S$-separable harmonic $\mc G_{[u]}^{\mc S}$-game. 
	\end{itemize}
\end{thm}
Here we state a few direct consequences of Theorem \ref{theo:main}. First, recall that, for every splitting $\mc S$ of $\mc G_{[u]}$, we have that $\mc G_{[u]}^{\mc S}\subseteq\mc G_{[u]}^{\triangle}$, so that Theorem \ref{theo:main} directly implies the following. 

\begin{cor}\label{coro:main}
	Every finite game $u\in\Gamma$ can be decomposed as  
	$$u=u_{\mathcal{N}}+u_{\mathcal{P}}+u_{\mathcal{H}}\,,$$ 
	where
	\begin{itemize}
		\item $u_{\mathcal{N}}$ is a non-strategic $\mc G_u$-game; 
		\item $u_{\mathcal{P}}$ is a normalized potential $\mc G_{[u]}^{\triangle}$-game; 
		\item $u_{\mathcal{H}}$ is a normalized harmonic $\mc G_{[u]}^{\triangle}$-game. 
	\end{itemize}
\end{cor}

Corollary \ref{coro:main} implies that $\mc G_{u_{\mc P}}$ and $\mc G_{u_{\mc H}}$ are subgraphs of $\mc G_{[u]}^{\triangle}$.
So, in general, the decomposition does not preserve graphicality, but in the graphs describing the interactions among players in the potential and harmonic component of a game $u$ there may be a direct influence only among players which belong to a common out-neighbourhood in $\mc G_{[u]}$.
Comparing Theorem \ref{theo:main} and Corollary \ref{coro:main} we see that the separated dependence of utility functions on neighbours allows to exclude the presence in the potential and harmonic component of direct interactions between players belonging to different groups. 
In fact, often the graph $\mc G^{\mc S}_{[u]}$ is much smaller than $\mc G^{\triangle}_{[u]}$. 
In particular, for the special case of pairwise-separable graphical games, Theorem \ref{theo:main} implies the following result.

\begin{cor}\label{prop:decomposition pairwise}
	Let $u\in\Gamma$ be a pairwise-separable graphical game on a graph $\mc G=(\mc V,\mc E)$, with utilities as in \eqref{pairwise-game}. Then, 
	$u=u^{\mc N}+u^{\mc P}+u^{\mc H}$ where 
	\begin{itemize}
		\item $u_{\mathcal{N}}$ is a non-strategic $\mc G$-game; 
		\item $u_{\mathcal{P}}$ is a pairwise-separable normalized potential $\mc G^{\leftrightarrow}$-game; 
		\item $u_{\mathcal{H}}$ is a pairwise-separable normalized harmonic $\mc G^{\leftrightarrow}$-game. 
	\end{itemize}
\end{cor}
This result shows that for pairwise-separable graphical games the decomposition in potential and harmonic part preserves the original graphical structure as it does not create any link between players that were not directly interacting in the original game. 

\subsection{Example}\label{example}
\begin{figure}
	\centering
	\includegraphics[width=0.4\linewidth]{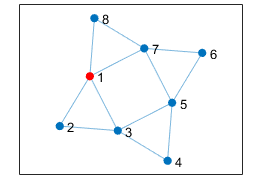}
	\caption{Interaction graph $\mc G$ for example \ref{example}.}
	\label{fig:example1}
\end{figure}

As an example of the results, consider the following $\mc G$-game. Eight players connected through the graph $\mc G$ as in Figure \ref{fig:example1}
decide of acquiring (action $1$) or not acquiring (action $0$) some good. All players but one are assumed to have an imitative behaviour, in that they just benefit from taking the same action as the majority of their neighbours. Accordingly, they play a majority game, with utility functions
\begin{equation*}
u_i(x) = \left| \left\lbrace j \in \mc N_i : x_j = x_i \right\rbrace \right| \quad i=2,\ldots,8 \,.
\end{equation*}
Player $1$ plays a so called ``public good game",
\begin{align*}
\begin{cases}
u_1(x)&{}= 1-c \quad \text{if } x_1=1\\
u_1(x)&{}= 1 \qquad \text{if }x_1=0, x_j=1 \text{ for some } j \in N(1)\\
u_1(x)&{}= 0 \qquad \text{if }x_1=0, x_j=0 \text{ for all } j \in N(1)\,,
\end{cases}
\end{align*}
i.e., she prefers to borrow the good from some of her neighbours rather than buying it and paying the cost $c$.
\begin{figure}
	\centering
	\includegraphics[width=0.8\linewidth]{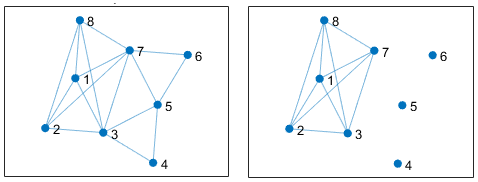}
	\caption{Minimal graphs 
		associated to the potential (left) and harmonic (right) component of the game in example \ref{example} (up to strategic equivalence)}
	\label{fig:example2}
\end{figure}
Figure \ref{fig:example2} shows the minimal graphs associated to the potential and harmonic components of the game. According to the theorem, they are subgraphs of $\mathcal{G}^{\triangle}$ and in this case they are proper subgraphs. Actually we see some additional feature of the decomposition. The game is not potential but is indeed a local perturbation of a potential game, the majority game: the locality of the perturbation is preserved by the decomposition. Indeed, we see that players which are far from the perturbation are only linked in the graph of the potential component and that the only additional edges with respect to $\mathcal{G}$ are between neighbours of player $1$.

\section{Conclusion}
We analysed the interaction between graphicality and strategic equivalence, proving the remarkable fact that there exists a minimal graph associated to a game up to strategic equivalence.
The core of our analysis has been the understanding of the interplay between graphicality and decomposition, yielding our major result which shows that the potential and harmonic components of a game $u$ are graphical games with respect to the undirected graph $\mc G^{\triangle}_{[u]}$. Current work includes the interpretation of the role of hidden strategic interactions and the application of these techniques to the theory of Markov Random Fields (in the spirit of \cite{Babichenko.Tamuz:2016}) and as a tool for studying Nash equilibria of perturbations of potential games.
\bibliography{bib}

\end{document}